# Black Feminist Musings on Algorithmic Oppression

Lelia Marie Hampton
EECS and CSAIL
MIT
Cambridge, MA, USA
lelia@mit.edu

## ABSTRACT

This paper unapologetically reflects on the critical role that Black feminism can and should play in abolishing algorithmic oppression. Positioning algorithmic oppression in the broader field of feminist science and technology studies, I draw upon feminist philosophical critiques of science and technology and discuss histories and continuities of scientific oppression against historically marginalized people. Moreover, I examine the concepts of invisibility and hypervisibility in oppressive technologies a lá the canonical double bind. Furthermore, I discuss what it means to call for diversity as a solution to algorithmic violence, and I critique dialectics of the fairness, accountability, and transparency community. I end by inviting you to envision and imagine the struggle to abolish algorithmic oppression by abolishing oppressive systems and shifting algorithmic development practices, including engaging our communities in scientific processes, centering marginalized communities in design, and consensual data and algorithmic practices.

## 1 Introduction

The following is a roadmap. This paper uses a theory of oppression to ground and extend algorithmic oppression. Algorithmic oppression is then situated through a Black feminist lens part of which entails highlighting the double bind of technology. To reconcile algorithmic oppression with respect to the fairness, accountability, and transparency community, I critique the language of the community. Lastly, I place algorithmic oppression in a broader conversation of feminist science, technology, and society studies to ground the discussion of ways forward through abolition and empowering marginalized communities.

## 2 A Theory of Algorithmic Oppression

First, we explore the theory of oppression from a Black feminist perspective and then we relate this back to what Dr. Safiya Noble calls *algorithmic oppression* [28].

### 2.1 Understanding Oppression

Oppression. *noun*. "A social system of barriers that operate institutionally and interpersonally to disempower people because of their gender, race, class, sexuality, ethnicity, religion, body size, ability, and/or nationality." [30]

Oppressive institutional systems and forces shape and confine oppressed people's lives in a way that is intentional, chronic, and *inescapable* [10]. Even more so, these systems operate interactively and inseparably as to "penalize motion in any direction" [10, 3, 5]. To be clear, oppression dictates that oppressors will inflict suffering upon the oppressed [10]. It follows that there is a group who constructs and maintains these hegemonic power structures as its existence serves their interests by confining, reducing, and immobilizing groups of people [10]. Those who benefit from oppression are young, thin, abled, Christian, affluent, cisheterosexual, white, men, etc. [22]. Thus, these systems of oppression are what bell hooks refers to as the white supremacist capitalist (cishetero)patriarchy [14, 15].

As Audre Lorde, the Black, lesbian, feminist, socialist, poet, mother of two including one boy and a member of an interracial couple, so eloquently points out, "there is no hierarchy of oppression" [24]. In other words, all oppression must be abolished, or oppression will never cease. This idea is especially important for those who face oppression at multiple fronts who do not have the luxury to battle oppression at one front [24].

## 2.2 Algorithmic Oppression

Algorithmic oppression solidifies the idea that institutions, structures, and systems of oppression are fundamental to the perpetuation of oppressive technology. While Noble's work focuses on people of color and women of color, I will extend this definition to all forms of oppressed people in this work (e.g. disabled, poor, queer, trans, etc.). Algorithmic oppression is a fitting point of departure for this work because dominant voices in the fairness conversations are reinforcing the power imbalances inherent in algorithms in the age of neoliberalism, and the consequent analysis isn't appropriate because it's not radical enough to address the issue at the root. In fact, the current discourse at present not only does not resolve any of the power polarities present in awry algorithms that are presented as "algorithmic bias", but it lacks the wherewithal to ever solve the problem technically if it cannot even admit that there is an irreformable problem societally. Moreover, the commonplace instances of technology going awry against oppressed people are not merely mistakes, but rather reverberations of existing global power structures [28]. Algorithmic oppression as a theoretical concept acknowledges that there are systems of oppression that cannot simply be reformed, and it does not only seek technical solutions to societal problems. Even more so, algorithmic oppression analyzes the ways that technology has violent impacts on marginalized people's lives, and in doing so it does not water down the impact to "discrimination" or "implicit bias" because doing so fundamentally invalidates the everlasting hell that oppressed people endure. And finally, algorithmic oppression cuts through the bullshit by unapologetically affirming marginalized people's experiences with oppressive technologies and prioritizing their perspectives in addressing them. Noble predicts that "…artificial intelligence will become a major human rights issue in the twenty-first century" [28], but I argue that it already has. Therefore, we must move toward a theory of algorithmic oppression because the context of technology cannot be divorced from global power structures [28].

Despite Safiya Noble releasing Algorithms of Oppression three years ago, the phrase *algorithmic oppression* has not gained much traction within the technology and ethics community, with many opting to use algorithmic bias to refer to unjust manifestations of algorithms which enforce and reinforce "societal bias". It seems that the word oppression is a polarizing one, and as Marilyn Frye puts it "The word 'oppression' is a strong word. It repels…" [10]. However, in a way, bias removes responsibility and makes it seem as if what is happening is "accidental" or a "side effect". Moreover, the use of "bias" obscures the very real social structures that are contributing to the output of these algorithms, creating a picture that in neoliberal fashion it is unintentional or not in bad faith rather than an intentional byproduct of oppressive institutions. That is, this claim purports the idea that well-meaning human beings are unconsciously making these tools, but as the following sections point out, that is not necessarily the case. Even more so, it downplays the severity with which algorithms can and do impact oppressed people. For example, there is nothing well-intentioned about predictive policing resulting in perpetual overpolicing of Black and Indigenous neighborhoods or using autonomous AI weapons to murder Black and Brown people and children.

To tie things back together, we cannot discuss algorithmic oppression without discussing systems of oppression because a struggle for liberation from algorithmic oppression also entails a struggle for liberation from all oppression as the two are inextricable. For example, one of the pillars and points of machine learning is "decision-making". But who is making decisions and for whom? Of course, it is often affluent abled cisheterosexual (white) men who often occupy other high positions on the social hierarchy. Even more so, there is an oligopoly of U.S. technology companies run by mostly affluent cisheterosexual white men and their accomplices who globally control technological infrastructure and make the products that billions of people around the world use to create, share, and access information. Consequently, they have a massive influence on the shaping of our worldview. Thus, algorithms are "loaded with power" [28].

## 2.3 Will more diversity in tech solve algorithmic oppression?

Will more diversity in tech solve algorithmic oppression? Not necessarily. As Ruha Benjamin points out: "...just having a more diverse team is an inadequate solution to discriminatory design practices that grow out of the interplay of racism and capitalism" [1]. In fact, it is very much a neoliberal concept. Ultimately, it shifts responsibility from 'our technologies are harming people' to 'BIPOC tokens have to fix it'. This practice is a way to mimic corporate social responsibility for branding purposes without materially changing the conditions of current BIPOC engineers, addressing the violent outcomes of technologies, and most importantly acknowledging responsibility for harming communities. Bringing Black tokens into capitalist profit-first tech companies does not necessarily fix issues in technology, especially when their voices and existence are discarded, and they are treated as an incompetent, undeserving, unworthy diversity hire who has nothing to contribute but a 0.0001% increase in the number of Black people at the company to help the company's reputation fare better while "diversity and inclusion" and "social responsibility" are trendy and profitable right now. The argument that diversity can solve this problem is immediately countered by the ousting of Timnit Gebru and April Christina Curley by Google in the middle of a global pandemic despite their passionate advocacy for more equity and massive contributions to the company. When we promote diversity as a solution to algorithmic oppression, we obscure the fact that there are power imbalances that are deeply embedded in societal systems and institutions. To expound, a) the Black persons' voice will not necessarily be valued or considered and b) a singular Black person on one team at a top tech company with a couple hundred Black employees out of tens of thousands of employees will not dislodge the racial inequity and violence inherent in the manifestations of these technologies. Even more so, putting Black people in a white masculinist space on their own can actually be extremely detrimental to their well-being as they will inevitably experience racism in one form or another in a predominantly white masculinist company built into the power structures of white supremacist capitalist cisheteropatriarchy.

## 3   Why a Black Feminist Lens?

In brief, radical Black feminism is the study and praxis of the end of all oppression everywhere [14, 15]. That is, the end of misogyny, misogynoir, cissexism, heterosexism, queerphobia/homophobia, ableism, Islamophobia, anti-Semitism, classism, capitalism, ageism, and so on (notice how I mention 13 forms of oppression and still had to specify there are more). Black feminism respects people's lived experiences and asserts people's right to speak for themselves and their needs in the abolition of their oppression—an important idea in reimagining technology after the abolition of algorithmic oppression. In fact, people's lived experiences are the center and start of theorizing, and Black feminist epistemology validates people's lived experiences as a way of knowing unlike white supremacy which cannot validate the oppressed's lived experiences if it wishes to perpetuate [3]. That is, if people do not understand en masse the severity of oppression associated with white supremacist capitalist cisheteropatriachy, then they will not resist this domination.

Consequently, A Black feminist lens is essential in approaching algorithmic oppression because it provides context that decenters the dominant lens (i.e. white supremacy, capitalism, cisheteropatriarchy, etc.) through which we understand oppressive algorithmic manifestations [28]. In fact, I would plainly state that algorithmic oppression cannot be overcome without integrating a Black feminist lens, a lens which empowers all peoples, a lá its stance that every form of oppression must be abolished. Unlike white male epistemology, which is placed at the center of the worldview, Black feminism validates ways of knowing outside of this centered worldview, resists suppression of oppressed people's knowledge production, and challenges white supremacist capitalist cisheteropatriarchal hegemony [3]. In this case, this knowledge produced by marginalized people with respect to Black feminism is in part the violent experiences of the oppressed at the hands of technology which I expand on in the following section. The Combahee River Collective highlights one of the reasons Black feminism is so powerful as an analytical tool and operational praxis: "If Black women were free, it would mean that everyone else would have to be free since our freedom would necessitate the destruction of all systems of oppression", drawing from the assertion that Black women can be subject to every form of domination [31]. That is, if we liberate (trans) Black women from algorithmic oppression, we liberate *everybody* from algorithmic oppression, hence the power of the Black feminist perspective.

I will take it one step further to say that in computing academic and industry circles there is a suppression of Black Feminist Thought, and as Patricia Hill Collins puts it: "This larger system of oppression works to suppress the ideas of Black women intellectuals and to protect elite white male interests and worldviews" [3]. Suppression of Black women intellectuals in technology is apparent in the treatment of Timnit Gebru by Google executives for her work on natural language models and unapologetically advocating for marginalized people. In part, suppression of Black feminist thought has gotten us into this algorithmic oppression nightmare. For instance, white men are not mothers of daughters who face racist and sexist violence, and therefore have not stopped to think about the pain and very real consequences of their technologies, i.e., being fetishized in a search engine, when we mothers know being fetishized perpetuates sexual violence against us and our daughters, as we are not seen to be humans but rather objects to suppress. To add to this point, suppression in this specific example is also due to the reality that sexism requires sexual violence to perpetuate [17]. Thus, sociotechnical design is not only embedded with oppression but also perpetuates it.

We cannot continue to use white supremacist capitalist cisheteropatriarchal ways of knowing to dismantle algorithmic oppression, "For the master's tools will never dismantle the master's house. They may allow us temporarily to beat him at his own game, but they will never enable us to bring about genuine change" [23]. That is, a reformist approach rooted in white male epistemology will not work. This idea is *reiterated* by the reality that something is obviously not working with the current mode of approaching algorithms and their violent outcomes because there is still algorithmic oppression at the hands of white supremacist capitalist cisheteropatriarchal AI tools. One might retort: "But the tools obviously work; look at all we've been able to accomplish"; okay, but if your machine learning model had 50% risk/loss/error, would you turn it in for your homework (deployment was originally used for this metaphor, but it doesn't work here because inaccurate ML models are frequently deployed to the detriment of marginalized people)? Then why are we allowing algorithmically oppressive tools to fail oppressed people for the sake of "science"... and profit?

## 4   Algorithmic Oppression through a Black Feminist Lens

What we mean by examining algorithmic oppression through a Black feminist lens is that we want to emphasize and critique the material outcomes of haywire technologies on oppressed peoples' lives. It is not simply a matter of awry algorithms that "accidentally discriminate". Canonical works on algorithmic oppression such as Gender Shades have been watered down to "bias" while misunderstanding the violence inherent in misogynoir, colorism, and misgendering Black women given the impact of the legacy of slavery on Black womanhood (i.e., that Black women are not women, or even human at all). In fact, technology should not gender or surveil anyone [20, 21]. In other words, these oppressive technologies have very real consequences on our lives as outlined below.

*[Trigger warning: mentions of violence against oppressed people]*
Please note that the following is not sugar coated as I should have not to exonerate the violence the oppressed face at the hands of AI made by oppressors nor placate the feelings of those who materially benefit from oppressive systems when our *lives* are at stake. I contend that placation is an absolution. Consequences of AI systems are literally the lives of Black and Brown people while it is the discussion of "fairness" to white people. As Black and Brown people, we have to worry about the police state weaponizing predictive policing (and please understand that there is no difference between the police state and police brutality as the goal of the police state is to brutalize, that is, "police" its citizens, those who are Black and Brown especially). Black and Brown folks overseas have to worry that military AI weapons will be used to murder them or their children (while they simply sleep in their home country). For trans folks, it is the violence of being misgendered, surveilled, targeted, and outed [20, 21]. For women, it is the acceleration and reinforcement of the pay gap which has even harsher material consequences for single-income mothers and women of color, i.e., filtering that shows higher-paying job ads to men more often than women [1]. For poor folks, it is being systematically denied healthcare, banks loans, and the same access to resources as affluent people [1, 9]. Another example of violence against poor folks is the steady incorporation of predictive analytics by U.S. social welfare agencies [1, 9]. For disabled people, it is healthcare companies weaponizing nonconsensually-collected health data to deny coverage in a society that already pushes them to the margins socially and economically. For Muslim folks, it is being stalked and systematically targeted by the federal government for simply existing and believing (while white U.S. citizens preach and praise first amendment "rights"). For Black and Brown mothers and our teenage daughters, it is our daughters being sexually fetishized in a search engine after centuries of sexual violence at the hands of the white supremacist cisheteropatriarchal settler colonial state [28]. Expounding on this violence, I invoke Audre Lorde's explanation of personhood and motherhood in an oppressive world, "But Black women and our children know the fabric of our lives is stitched with violence and with hatred, that there is no rest. We do not deal with it only on the picket lines, or in dark midnight alleys, or in the places where we dare to verbalize our resistance. For us, increasingly, violence weaves through the daily tissues of our living" [22]. For Black folks, it is searching ourselves in a search engine and seeing "gorilla" or "ape" after centuries of violently racist epithets used to deny us personhood and humanity [1, 28]. All of these are very much violent and exhausting for oppressed people, and they take a toll on our physical and mental health. To be called anything less than violent is to severely miscomprehend the severity with which algorithms and artificial intelligence gravely impact the lives of the oppressed. Even more so, the question of intention is less important than the actual outcomes upon people's lives.

To be clear, we can never have "equality", "fairness", or any other neoliberal buzzwords in AI until we abolish the white supremacist capitalist ableist ageist cisheteropatriarchy. Even more so, we have a field (computer science) and its subfield (artificial intelligence) which have systematically kept out Black and Brown folks and then "inadvertently" created "biased" projects. Then we have the other case, in which affluent cisheterosexual white men and their accomplices very much intentionally create AI technologies to weaponize against BIPOC, poor people, transgender people, etc. And of course, predictive policing and "criminal justice" algorithms also fall into this category. In fact, companies have taken people's photos off social media with billions of users, collecting extremely massive datasets for facial recognition technologies for (ab)use by law enforcement, largely without knowledge and wholly without consent.

Of course, when one does not experience oppression in the forms of racism, sexism, transphobia/transmisogyny/transmisogynoir, classism, ableism, Islamophobia, queerphobia/homophobia, and so on, then one would not consider the fact that technologies could be weaponized against others. Even more so, some simply do not care and blatantly and intentionally disregard others' lives. This violence leads to a couple of questions. How do we decolonize technologies in a settler colonial state that continues to abuse BIPOC citizens through the prison industrial complex and military industrial complex? How do we decolonize AI if the world is not even decolonized? Consequently, if we wish to live in a more just world inside and outside of technology, we must not only abolish algorithmic oppression, but all oppression.

## 4.1 Invisibility versus Hypervisibility in (AI) Algorithms

Algorithms have at once invisibilized and hypervisibilized the oppressed. This paradigm is where context matters and a "one size fits all" solution does not work and will inevitably fail; in fact, we must push aside this black/white thinking inherent in white supremacist cisheteropatriarchal ways of knowing.

Let's use the "Gender Shades" study as an example of invisibilization [2]. On the one hand, you have a tool that misgenders and feeds into the idea of Black women not being women vis-à-vis white supremacy, the legacies of chattel slavery, and the construction of white womanhood in juxtaposition to Black womanhood, thereby enacting misgynoir. Being misgendered and experiencing racism, sexism, and colorism at the hands of a machine is detrimental to one's psyche and could contribute to what The Combahee River Collective calls "The psychological toll of being a Black woman..." [31]. Even more so, invisibilization could have harmful socioeconomic impacts for marginalized groups. For instance, being invisibilized could harm BIPOC who use the Internet to promote their products and services as it negatively impacts their economic livelihood on top of being economically suppressed historically. For example, I highlight in a Medium post that Twitter's preview feature only showing light skin sabotaged a Black artist's income since their work featured Black subjects, and thus did not show these subjects in the preview feature [41]. As a final example of invisibility, an algorithm for a healthcare management program, which is designed to provide additional benefits to high risk patients and makes decisions for 70 million people, systematically denies these benefits to Black patients, despite the fact that they pay more out of pocket on average [42]. Moreover, the algorithm gives high risk Black patients with more chronic illnesses the same risk score as white patients, thus contributing to a long historical continuum of medical racism [42].

On the other hand, there's the case of hypervisibility such as tools that automate oppression (i.e. algorithms made to systematically target disabled people and/or poor people and deny them resources), for example the use of predictive analytics in the U.S. welfare system [9]. Even more so, there are algorithms, such as predictive policing algorithms, which are being used to intentionally violently oppress and specifically target BIPOC communities and feed us to the prison industrial complex. Do I, a Black person, really want a police state surveillance tool to recognize my face in order to further criminalize and marginalize me? *Hell* **NO**! And I'm sure my Indigenous and Black/Brown Muslim comrades feel the same sentiments. Moreover, Os Keyes also provides insight into the inherent violence in trans people being hypervisibilized with respect to data collection in "Counting the Countless" [20]. Companies, the government, medical professionals, etc. use their power to enact administrative violence to enable control and surveillance of trans people. That is, trying to live in an affirmative way by openly and proudly being trans results in even more corporate and state surveillance (which are inextricable at this point).

With respect to invisibility and hypervisibility, oppressed people are damned if we do, and damned if we don't. That is, there is always a violent outcome. This phenomenon is known as the double bind, which is present in both race and gender theory and I suppose all theories of oppression, whether formally or informally. The double bind demonstrates that white supremacist cisheteropatriarchy transforms to fit its purpose, that is, a violent, inescapable oppression that solely materially benefits cisheterosexual affluent white men. To tie things back together, the concept of invisibility versus hypervisibility relates to the inescapability and chronicity of oppression, and in this case algorithmic oppression.

## 5  Dialectics of The Language of the AI and Ethics Community

I would like to take a moment to examine the use of "fairness", "accountability", and "transparency". To be clear, this is a critique rather than a request to alter the language. Indeed, it is important for us to reflect on our language. After all, our work impacts communities of people beyond us and our research labs. In fact, our work is (or at least should be) important in eliminating harm vis-à-vis algorithmic oppression. I agree with Ruha Benjamin that some of our approaches to algorithmic oppression are "technological fixes that encode inequity in a different form" [1]. In fact, this unfortunate approach is present in most, if not all, of the literature on algorithmic fairness approaches to recidivism (as one example out of many). As a Black person, it is even more disturbing to see cisheterosexual affluent abled thin mostly white male academics argue over whether risk assessment tools in the punitive system should use the "disparate impact" equation or the "demographic parity" equation. While Black and Indigenous people waste away in 3 foot by 3 foot cages and are denied basic human necessities and rights, these powerful mostly white male academics are arguing about which equations should be used to throw the niggers in prison in the comfort of their lavish homes and well-endowed labs at prestigious universities in neighborhoods that Black and Indigenous people are not even allowed in thanks to "de facto" segregation and policing. Ain't that about a bitch?

As of late, there is much talk of including communities outside of the computing one to help us address algorithmic fairness [1, 4, 25, 37]. However, the idea of "fairness" is not only confusing but it also does not intuitively tell us what it means to pursue algorithmic fairness. What is "fairness", and who defines it? Well, there is no agreed upon definition of algorithmic fairness that can be written in a dictionary yet (or even searched up online; trust me I've tried); in fact, one source cites twenty-one linguistic definitions of algorithmic fairness [34]. One academic argues that we need not have a singular definition of fairness, but I believe we need not have the concept of fairness at all if it refuses to acknowledge that technologies reverberating the sociohistorical subjugation of oppressed people are irredeemable. How can one be "fair" in an unjust society? After all, if there is a concept of limited resources with respect to capitalism, then it follows that there is a group of people who will not have access to these resources who will inevitably be marginalized to achieve this hierarchy of resources (and thus we have racial capitalism). Interestingly, there have been a number of equations to come out of the algorithmic fairness literature which seek to use these equations to examine algorithmic output's treatment of different populations [35]. But even this isn't seemingly intentional in its desired outcome or even properly contextualized. Say for instance, in a make-believe world, white men were being targeted by the police state en masse with some algorithm, and the mathematical notions for disparate impact (e.g. outcomes disproportionately hurt white men) was used to discern if people of color were experiencing it at the same rate. Do you think I or any other person of color want the algorithm to be "fair", distribute this burden, and target us too? *Hell* **NO!** In fact, I do not want the police state to target anybody or hand any of us over to the prison industrial complex to be abused as slave labor at 10 cents an hour. And *that* is the importance of contextualizing algorithmic fairness and intentionally ensuring algorithms do not harm people—another reason Black feminism is a necessary lens here.

Even more so, until white researchers engage in radical theory and praxis, then they are not only incapable of advancing "fairness and accountability", but also complicit in the violence perpetuated by AI against oppressed people. While academics make up equations, Western imperialist countries use autonomous weapons to murder people, and mostly white male judges use recidivism algorithms for death penalty cases. How do you "fairly" administer arbitrary death by the state for being Black or Brown? Ultimately, we cannot abolish algorithmic oppression without abolishing systemic oppression as the two are indivisible. To be honest, I am unsure that algorithmic fairness as a discipline can be redeemed since it is based in reformist ideology which serves to perpetuate algorithmic oppression while purporting to offer solutions which are all in all half assed because for the 10000th time they do not consider that some of these technologies should not even exist. The example with autonomous weapons and recidivism algorithms sadly drives this point home. While white men were arguing amongst themselves from their lavish homes and faculty positions at Ivy plus institutions about which equation should be used [39], poor Black and Brown people were rotting in cages and being told to kick rocks. In fact, the whole debate surrounding which equations should be used to analyze recidivism algorithm for "fairness" serves as a case study that algorithmic fairness does not have the wherewithal to advance algorithmic justice if it chooses to legitimize risk assessment algorithms and the criminal justice system at the expense of Black and Brown lives and communities who are ravaged by the police state and prison industrial complex.

Let's examine the use of "transparency". Transparency (i.e. open source) is a beautiful part about the computing community and shows that we do actually care about justice to some extent. But is it always beneficial in addressing algorithmic oppression? For instance, what if we make facial recognition technology used by U.S. immigration services open source, "interpretable", or "explainable"? What benefits would this have to parents separated from their children by immigration officers who don't value their personhood or their life? That is to say just because machine learning is transparent doesn't mean it's not oppressive. I suppose some might argue that it is necessary for criminal defendants that risk assessment algorithms be transparent to argue against them in court. However, this notion supports the idea that punitive systems will act rationally when shown counterevidence, and history has shown that these judges don't give a damn. They frequently perpetuate epistemic injustice (i.e. oppressors ignoring the testimonial of oppressed people). For example, judges have ruled against death penalty defendants' requests to push back execution in light of new evidence and have invalidated statistics demonstrating that Black people occupy most of death row. The punitive system in the U.S. is in place to incarcerate and surveil Black and Indigenous people en masse, and that is exactly the function it will reserve whether or not an algorithm is "transparent". That is not to say that naming and shaming isn't important and doesn't have its place in addressing algorithmic oppression, but by itself it cannot transform these power structures or moralize oppressors beyond feigned press releases claiming to care about whatever issue was highlighted. Therefore, it is important to be specific and intentional not only with our words, but also how we intend our work to be used to advance algorithmic justice—if we intend it to be used that way at all.

Something else I would like to critique is the tendency to call us "protected groups" in the algorithmic fairness literature when we are in fact not protected nor does it materially transform our circumstances, so it is in essence a meaningless gesture although well-intentioned. Who is protected? Certainly not us in this world we live. Protected by whom? Certainly not the militaristic police state nor the settler colonial state. Oppressed would be a more fitting word here, but again, as Frye points out the word oppression "repels" [10]. This brings me back to the use of "algorithmic bias" rather than "algorithmic oppression" in the technology and ethics community. As stated, bias implies lack of intention and less severe outcomes than what is actually occurring. As I have pointed out, there are companies who will make technologies at the expense of other people's lives for the sake of profit.

How do you hold oppressive technology accountable in a white supremacist capitalist ageist ableist cisheteropatriachal society where violence is dealt to the oppressed with impunity? That is to say, we must really think about the words we use very carefully as words have as much meaning as algorithms do in our lives. If we instead choose to be thoughtless with our words, we run the risk of never solving the actual problem at hand. Thus, we must continually ask "are we solving the symptoms or the problem?" and "are we getting to the root of the problem of algorithmic oppression?".

## 6 Situating Algorithmic Oppression in the Broader Purview of Feminist Science and Technology Studies

*"Every tool is a weapon if you hold it right." — Ani Difranco [8]*

If we place AI in the broader context of the history of science, we can understand the ways in which AI is continuing a long history of abusing poor Black and Brown folks, queer and trans folks, women and femmes, disabled people, and more in the name of science. We must always keep in mind that "Technologies also have genealogies and histories; their pasts bear legacies that ought not be forgotten, even if present-day usages, such as in 'hacks', differ from their provenance" [19]. Well-known examples include intentional refusal to provide treatment for Black men and their sexual partners for *decades* in the Tuskegee syphilis trials, the mutilation and murder of enslaved Black women and infants to establish the field of gynecology, and the exploitation of a terminally ill Henrietta Lacks whose cells have bolstered billions of dollars for pharmaceutical companies. Save some descendants of Tuskegee victims, none of them ever received compensation nor did their families, but their bodies, health, and well-being were deliberately harmed to benefit those who benefit from capitalist white male supremacy. This exploitation can also be paralleled with the ways queer and trans folks have been abused in science (i.e., castration for cis gay men to prevent sexual attraction toward other men or electric shock therapy for "conversion"). Science continues to gain knowledge, for the benefit of affluent cisheterosexual white people, at the expense of poor BIPOC/BIWOC, queer and trans folks, disabled folks, and so on. These infamous examples demonstrate that science has been used to abuse and exploit oppressed people in the service of systems of oppression, including racism, sexism, heterosexism, and classism [13]. The legacy of using disempowered people as nonconsensual and vulnerable test subjects continues in machine learning today. In fact, a Google contractor paid homeless people in Atlanta, a majority Black city, $5 to use their pictures in a facial recognition database. In addition, one study found that the U.S. government is using "images of children who have been exploited for child pornography; U.S. visa applicants, especially those from Mexico; and people who have been arrested and are now deceased" [40]. We must be constantly critical of the ways in which technologies are harming Black and Brown folks and marginalized people in general, and therefore constantly evolve our ethical practices to reflect our current realities. As an aside, one ethical practice that should be without a doubt implemented is a) no surveillance of citizens at all, especially using AI technologies, b) no technologies that throw marginalized people to the carceral or police state, and c) no autonomous AI weapons.

Feminist science and technology studies encourages us to ask: "Who gets to create what, for whose benefit, at whose expense?" [30]. Ruth Hubbard, who provides an insider's perspective in feminist science and technology studies as a biologist, contends that science is not made in a vacuum or outside of social systems, that is, "Every fact has a factor, a maker" [18]. Hubbard also encourages us to think about what social groups are allowed to make scientific facts and which are not because "whoever gets to define what counts as a scientific problem also gets a powerful role in shaping the picture of the world that results from scientific research" [18]. Even more so, Hubbard posits that as scientists "...we must follow the rules of the scientific community and go about our fact making in professionally sanctioned ways. We must submit new facts to review by colleagues and be willing to share them with qualified strangers by writing and speaking about them" [18]. To be clear, "the rules of the scientific community" serve as a form of gatekeeping which usually police marginalized people's ability to contribute to scientific knowledge production and push us out. In fact, those who decide who receive faculty or research scientist positions as well as those who decide whose "proposal has enough merit to be financed" are also gatekeepers, and thereby play a critical role in scientific knowledge production [18]. Note that this group of gatekeepers tends to be rather small, namely, "by the chosen for the chosen" [18]. While science is often perceived as "the absolute truth" or "value-free", scientists' findings are shaped by their social context, that is, the dominant ideologies (i.e. racism, sexism, heterosexism, etc.) which impact the scientific questions and problems they pursue, the methods they employ, and the conclusions at which they arrive [18, 30]. Consequently, we must critique in a principled manner the production of scientific knowledge, examining researchers' situated knowledge (i.e. the idea that "knowledge is always produced from a specific disciplinary, ideological, and social location"), their institutions, their funding sources, their methods, the people who will benefit from the research, and the people who will potentially be harmed by the research [30].

Despite evidence of science being socially constructed, we must not do away with science, but rather find ways forward by producing a situated knowledge and "avoid occupying the omniscient position of an all-knowing god" [30, 38]. In fact, feminist science and technology studies encourages us to envision "how to create the kinds of societies in which the dominant institutions of knowledge production are no longer so complicitous in benefiting the few to the detriment of the many" [18]. Donna Haraway encourages us to "be critical of why certain technologies receive more funding and attention than others and how these technologies perpetuate global inequalities", and I think this question is especially important given a) mounting cases of algorithmic oppression at the hands of AI and b) the recent announcement by the National Science Foundation that they will be giving special attention to artificial intelligence and quantum computing in the next Graduate Research Fellowship Program cycle [30, 38]. To expound, the U.S. government has an incentive to fund AI research since it is in a battle to be the head of the AI superpowers, and AI can be used for imperialist purposes such as autonomous weapons. Even more disturbing as Sandra Harding highlights is the "Western thought" that scientific abuses against oppressed people are necessary and should not deter "the inevitable benefits of greater knowledge and or greater choices" [13]. Moreover, technologies, applied sciences, and scientific theories have historically been leveraged to shift control from the oppressed "to those who exercise power in the dominant class, race, and culture" [13]. Essentially, the sciences have helped to advance systems of oppression [13]. That is, while benefits accrue for affluent cisheterosexual white men, abuses and disadvantages accumulate for BIPOC, women, trans/queer people, disabled people, elders, and other marginalized groups [13].

Another interesting idea to consider here is "masculinist" algorithms, or algorithms and technologies that enforce and reinforce cisheteropatriarchy a là their creators: "Many big tech companies are located in Silicon Valley, a place overpopulated by male tech workers. This, in turn, shapes its masculine work culture and the products and mobile apps they produce" [30]. Notably, machine learning is using data that is shaped by society, which is also shaping the output of the algorithm, entailing that the machine learning model is socially constructed as a result. An example of a socially constructed machine learning model is predictive policing algorithms which are shaped by racist/sexist developers, racist/sexist data, and racist/sexist narratives. Note that I point out sexism here to center and give voice to BIPOC who are women and women-assumed (e.g. nonbinary folks perceived as assigned female at birth) who are being targeted by the police state, prison industrial complex, and surveillance state, but frequently are invisibilized in these conversations.

Although science and technology can be detrimental to marginalized groups, there are roles for computing in social change [43]. I present an example of empowered Black women engineers who also empower their community through technology. Five Ugandan women at Makerere University, Margaret Nanyombi, Jackline Namanda, Esther Ndagire, Pauline Nairuba, and Bridget Mendoza, founded Team Code Gurus. They are the creators of the Her Health BVKit, a technology that determines the likelihood of having bacterial vaginosis and then directs high likelihood users to the nearest medical resources [19]. They planned to market the technology to nongovernmental organizations that travel to women who do not have health facilities [19]. Clare Jen describes this as an example of feminist hacktivism, or "oppositional endeavors that infiltrate and/or remake dominant technoscientific assemblages of people, objects, governance, and relations—all in order to meet on-the-ground needs of women and children. These infiltrations and remixes work toward visions of social justice in their local and global particularities" [19]. I bring up this example because it demonstrates the power of making not-for-profit technologies and being empowered as an engineer to solve problems that are important to oppressed people and our communities. In fact, this example juxtaposes the capitalist technology industry that engineers technologies to make profit at oppressed people's expense (as well as the idea pushed by Silicon Valley that Black people are not technically equipped to solve problems). Indeed, everybody should strive for engineering for social good, that is engineering which benefits all people and not the few, or not strive for engineering at all, if it comes at the cost of literal lives of oppressed folks. That is also to say, if technology benefits the majority, but harms a few, who will more than likely be oppressed, then we mustn't allow these technologies either.

As Clare Jen argues: "Technologies are not simply either 'good' or 'bad', yet they are not neutral" [19]. That is, there is (probably) nothing inherently evil about a neural network by itself, but when Western governments use them in autonomous weapons then it does in fact become evil. This notion is important to acknowledge lest we succumb to the all or nothing thinking inherent in white male supremacist epistemology. In other words, we must appreciate nuance and the way it shapes global geopolitical power contexts because "context matters when appraising a technology's potential to help or harm" [19].

## 7 Imagining and Reimagining Abolition of Algorithmic Oppression

*"Calls for abolition are never simply about bringing harmful systems to an end but also about envisioning new ones" — Ruha Benjamin [1]*

At present, we have a handful of people who influence and decide the scientific problems which impact the masses and not always for the betterment of the broader society. In addition to contextualizing algorithmic oppression in the realm of systems of oppression, we must also approach algorithmic oppression from the vantage point that "concerns surrounding algorithmic decision making and algorithmic injustice require fundamental rethinking above and beyond technical solutions" [12, 36]. Often, these technical solutions do not consider the context despite Black women [1, 12] repeatedly screaming at the top of our lungs that technical solutions mean nothing without the context, are rooted in reformism/neoliberalism, and do not solve any of the issues they purport to resolve. In some cases, there is not a technical solution at all. Some shit simply needs to be abolished. You cannot reform the sociotechnical afterlife of chattel slavery and colonialism. You must abolish the sociotechnical tools that are legacies of these institutions.

Moreover, we must consider the sociopolitical implications of our current scientific processes (i.e. professionally, methodologically, etc.) and modes of communicating our research to the public [18]. As we seek a way forward, we must acknowledge that "...in order to reform the sciences, as well as the philosophy, history, and social studies of science, there… have to be broad transformations of both science and society" [13]. Given the blatant disregard for (oppressed) human life, we must involve broader communities outside of industry and academia in developing technologies because "survival is not an academic skill" [23]. As we involve broader communities, we must ensure "A wide range of people...have access to scientific facts and to understanding and using them" [18]. Of course, this inspires the question: "How can we use for emancipatory ends those sciences that are apparently so intimately involved in Western, bourgeois, and masculine projects?" [13]. We have already demonstrated the case of Team Code Guru's "feminist hacktivism" [19]. Team Code Guru also shows us why we should fund BIPOC, women, queer/trans, disabled, and other marginalized scientists to carry out our own research agendas, so we can have some semblance of agency in scientific knowledge production, especially one that centers us in a field that uses our marginalized status to profit. Technologies created by our communities for our communities, especially for liberatory ends, relates to the principles of design justice, particularly "design to sustain, heal, and empower our communities, as well as to seek liberation from exploitative and oppressive systems" and "sustainable, community-led and - controlled outcomes" [37]. For this reason, I co-founded Blackathon, a hackathon by Black people, for Black people that allows us to create liberatory technologies for our communities while centering participants well-being and divesting from capitalist constraints.

Ultimately, science must be made by the people, for the people. Ruth Hubbard echoes these sentiments: "We must broaden the base of experience and knowledge on which scientists draw by insisting that a wider range of people must be able to do science and do it in a greater variety of ways and institutions. We must also insist that scientists provide information and understanding that can be useful to and used by different kinds of people" [18]. As Audre Lorde teaches us, "Without community there is no liberation, only the most vulnerable and temporary armistice between an individual and her oppression" [23]. Our communities must be empowered to engage in both participatory research and citizen science. Involving communities in science builds public accountability into scientific knowledge production, allowing important scientific questions to be generated by collective and consensual processes [18]. In engaging the broader communities in science, we must not assume we know more than them about their social context or scientific/engineering needs, and we must not make the elitist mistake of assuming they do not have the capacity to understand science. With relation to algorithmic oppression, "Relational ethics asks that for any solution that we seek, the starting point be the individuals and groups that are impacted the most" [36]. This idea relates to one of the principles of design justice: centering the voices of those who are directly impacted by the outcomes of the design process [37]. Empowering the masses to engage in technological development in numerous ways is a "greater investment in socially just imaginaries" [1].

We have a long way to go in engaging communities, especially since a recent survey showed that countries in Africa, South and Central America and Central Asia are underrepresented in the AI ethics debate even though they experience ethics dumping (e.g. the Cambridge Analytica scandal) and digital colonialism [25]. However, we are making some small steps in the right direction, such as the emerging field of participatory machine learning for example (although folks are still hesitant to do participatory research in science and engineering). Similarly to human-computer interaction and user-centered design, participatory machine learning draws upon critical technical practices: "Critical technical practices take a middle ground between the technical work of developing new AI algorithms and the reflexive work of criticism that uncovers hidden assumptions and alternative ways of working" [25]. Importantly, human-computer interaction and user-centered design ask for *consent* before developing technologies for populations. All in all, human-centered artificial intelligence using design justice and critical technical practices is a way forward, but again we need to ensure people in every geopolitical context have access. Even more so, we need to take care to not replicate the social inequities currently occurring in the field of human-computer interaction.

Notably, as a direct result, encouraging user design and feedback indirectly means we must rethink the way we view publications and research output in computer science. In particular, folks may not want to add anything to the research process that increases labor and time to completion if it impacts the publication process (and thus tenure and academic competitiveness). For instance, qualitative analysis is labor intensive, and machine learning moves at a fast pace and everybody wants to beat everybody else to the punch. As one example to counteract hesitation, the Computer Supported Cooperative Work conference has created a great solution here by accepting papers quarterly, and no, this has not made the conference any less "prestigious". By the same token, we must understand the elitism, and thereby the racism, sexism, classism, and ableism, perpetuated when we make obscene amounts of publications and "prestige" of the venue one of our ways to evaluate research competence and notoriety. Because again, (white) affluent abled cisheterosexual men by and large determine what scientific problems are worthy and not worthy, hence it follows that BIPOC, women, disabled folks, and queer/trans folks are being pushed out of the ability to advance their careers and science through this avenue in numerous ways.

An important thought to keep in mind is that it is not only irresponsible to force our ideas of what communities need, but also violent, and therefore with respect to design justice: "Before seeking new design solutions, we look for what is already working at the community level" and "We honor and uplift traditional, indigenous, and local knowledge and practices" [37]. This notion is especially relevant when we consider that white supremacist capitalist cisheteropatriarchy imposes hegemony on the oppressed in the way of a savior complex. We must emphasize an importance of including *all* communities, and the voices and ideas of marginalized people must be centered as we are the first and hardest hit by algorithmic oppression. Ruha Benjamin provides some insight: "By deliberately cultivating a solidaristic approach to design, we need to consider that technology might be working just fine for some of us (now) but could harm or exclude others and that, even when the stakes seem trivial, a visionary ethos requires looking down the road to where things might be headed. **We're next**." [1].

Sasha Costanza-Chock suggests a deviation from universalist design since it takes a utilitarian approach (i.e., benefit the majority at the expense of the minority) that harms marginalized people, especially those who face more than one institution of oppression. That is, we must understand that for technology there is no "one size fits all" approach. For instance, when reading a PDF file, a visually impaired person might need accommodations that I do not need. Another relevant example that Sasha Costanza-Chock recounts is that millimeter wave scanners at the airport reproduce and violently enforce the gender binary vis-á-vis the Transportation Security Administration (TSA) by targeting trans people who do not fit neatly into cisnormative categories.

Community education and engagement is paramount in our fight against algorithmic oppression [1]. We will solve nothing without involving our communities, and we must take care to ensure we do not impose elitist ideas of who can and cannot do science and engineering. Part of the approach to address algorithmic oppression is to organize in collective defiance to these coded inequities as a community and in our communities [1]. In doing so, we must remember that "...not everyone is in an equal position to refuse, owing to existing forms of stratification" [1, 20]. However, this barrier/suppression does not mean we speak on behalf of communities or engineer on their behalf without their consent and feedback but rather that we "*Move slower and empower people*" [1].

A lovely example of oppressed folks engineering liberatory solutions is Dr. Kourtney Zeigler, a Black trans tech developer and co-founder of Appolition, a portmanteau of abolition and application, which is an app that, according to Dr. Zeigler, "converts your daily change into bail money to free black people". Ruha Benjamin praises Zeigler and Appolition, citing that "Appolition is a technology with an emancipatory ethos, a tool of solidarity that directs resources to getting people literally free" [1]. To this end, nobody is more invested in liberatory engineering than our own communities, and we as members of our communities have the expertise (despite what white supremacy and cisheteropatriarchy say) to design and implement technologies for ourselves. As one of the design justice principles teaches us: "We believe that everyone is an expert based on their own lived experience, and that we all have unique and brilliant contributions to bring to a design process" [37]. Ruha Benjamin elicits a call to action on this basis: "Let us shift, then, from a technology as an outcome to toolmaking as a practice, so as to consider the many different types of tools needed to resist coded inequity, to build solidarity, and to engender liberation" [1]. Recalling our discussion about how context matters, Ruha Benjamin argues: "Justice, in this sense, is not a static value but an ongoing methodology that can and should be incorporated into tech design" [1]. We need to encourage more collectives and community organizing around algorithmic oppression and support those that already exist such as Data for Black Lives and The Carceral Tech Resistance Network.

We must not impose a (Western/white) savior complex in our pursuit of abolishing algorithmic oppression, creating emancipatory technologies that liberate people, and envisioning a future in technology that does not harm people. We have to appreciate that the people who best know the technologies they need are those people themselves. Even more so, only solidarity in the struggle for abolition with our communities can liberate us.

## 8   Closing Thoughts

*"In our world, divide and conquer must become define and empower" — Audre Lorde  [23]*

As evidenced, any time a tool perpetuates violence, it is a reverberation of white supremacist capitalist cisheteropatriarchy, and hence cannot be divorced from this system. Thus, we cannot continue to discuss "fairness and accountability in AI" without also discussing these systems of oppression. Even more so, until white (male) researchers engage in radical theory and praxis, then they are not only incapable of advancing "fairness and accountability", but also complicit in the violence perpetuated by AI against oppressed people. Ultimately, we cannot abolish algorithmic oppression without abolishing systemic oppression as the two are indivisible.

Simultaneously, we must reimagine technology in a way that benefits us collectively rather than the oppressing class. We must center our communities in the design of science and technology a lá design justice. We must empower communities to engineer on their own behalf without imposing the (white) savior complex on them, assuming they do not have the wherewithal to form or address their own scientific problems, and assuming we know what is best for them.

## ACKNOWLEDGMENTS

Thank you to the anonymous reviewers for your invaluable feedback. Thank you to Rachael McLaughlin and Ananda Griffin for providing invaluable feedback and examining my arguments. Thank you to Lebert Lester III for our insightful conversations and his support. Thank you to Dr. Kristen McHenry, Professor Beverly Guy-Sheftall, the Spelman College Women's Research and Resource Center, and the Spelman College African Diaspora and the World Program for not only instilling me with the theory to understand myself and my world but also encouraging praxis. Lastly and most importantly, I thank the ancestors for continually guiding me and assisting me. Without y'all, I would not be here.

## REFERENCES

[1] Ruha Benjamin. 2019. *Race After Technology: Abolitionist Tools for the New Jim Code*. Wiley, New York, NY.
[2] Joy Buolamwini and Timnit Gebru. 2018. Gender shades: Intersectional accuracy disparities in commercial gender classification. In *ACM Conference on Fairness, Accountability and Transparency*, 77-91
[3] Patricia Hill Collins. 2002. *Black Feminist Thought: Knowledge, Consciousness, and the Politics of Empowerment*. Routledge, New York, NY.
[4] Costanza-Chock, Sasha. 2018. Design justice, AI, and escape from the matrix of domination. *Journal of Design and Science (JoDS)*.
[5] Kimberlé Crenshaw. 1990. Mapping the margins: Intersectionality, identity politics, and violence against women of color. *Stan. L. Rev.* 43, (1990).
[6] Angela Davis. 1981. Reflections on the black woman's role in the community of slaves. *The Black Scholar* 12, 6 (1981), 2-15.
[7] Ian Editor (Ed.). 2007. *The title of book one* (1st. ed.). The name of the series one, Vol. 9. University of Chicago Press, Chicago. DOI:https://doi.org/10.1007/3-540-09237-4.
[8] Ani Difranco. 1993. My I.Q. *Puddle Dive*. Righteous Babe, Buffalo, NY.
[9] Virginia Eubanks. 2018. *Automating Inequality: How High-Tech Tools Profile, Police, and Punish the Poor*. St. Martin's Press, London, UK.
[10] Marilyn Frye. 1983. *The Politics of Reality*. Crossing Press, Freedom, CA.
[11] Leela Gandhi. 1998. Postcolonialism and feminism. *Writing Women and Space: Colonial and Post-colonial Geographies* (1998), 81-101.

Black Feminist Musings on Algorithmic Oppression				FAccT'21, March 3-10, 2021, Virtual Event, Canada[12] Timnit Gebru. 2019. Oxford handbook on AI ethics book chapter on race and gender. arXiv:1908.06165. Retrieved from https://arxiv.org/abs/1908.06165

[13] Sandra Harding. Feminism confronts the sciences: Reform and transformation. *Whose Science* (1991), 19-50.

[14] bell hooks. 2000. *Feminism is for Everybody: Passionate Politics*. Pluto Press, London, UK.

[15] bell hooks. 2000. *Feminist Theory: From Margin to Center*. Pluto Press, London, UK.

[16] bell hooks. 1990. Postmodern blackness. *Postmodern Culture* 1, 1 (1990).

[17] bell hooks. 1993. Seduced by violence no more.

[18] Ruth Hubbard. 2000. Fact making and feminism. *Gender and Social Life* (2000).

[19] Clare Jen. 2017. Feminist hactivisms: Countering technophilia and fictional promises.

[20] Os Keyes. Counting the countless: Why data science is a profound threat for queer people. *Real Life* 2 (2019).

[21] Os Keyes. 2018. The Misgendering Machines: Trans/HCI Implications of Automatic Gender Recognition. *Proc. ACM Hum.-Comput. Interact. 2,* CSCW, Article 88 (November 2018), 22 pages. DOI:https://doi.org/10.1145/3274357

[22] Audre Lorde. 1980. Age, race, class, and sex: Women redefining difference.

[23] Audre Lorde. 2018. *The Master's Tools Will Never Dismantle the Master's House*. Penguin, London, UK.

[24] Audre Lorde. There is no hierarchy of oppressions. *Bulletin: Homophobia and education* 14.3/4 (1983): 9.

[25] Shakir Mohamed, Marie-Therese Png, and William Isaac. 2020. Decolonial AI: Decolonial theory as sociotechnical foresight in artificial intelligence. *Philosophy & Technology* (2020): 1-26.

[26] Toni Morrison. A Humanist View. A Conference on Black Studies at Portland State University, 30 May 1975, Portland State University, Portland, Oregon. Keynote Speech.

[27] NAACP. Criminal Justice Fact Sheet. https://www.naacp.org/criminal-justice-fact-sheet/

[28] Safiya Umoja Noble. 2018. *Algorithms of Oppression: How Search Engines Reinforce Racism*. NYU Press.

[29] Cathy O'Neil. 2016. *Weapons of Math Destruction: How Big Data Increases Inequality and Threatens Democracy*. Broadway Books, New York, NY.

[30] L. Ayu Saraswati, Barbara L. Shaw, and Heather Rellihan. 2018. *Introduction to Women's, Gender & Sexuality Studies: Interdisciplinary and Intersectional Approaches*. Oxford University Press, Oxford, UK.

[31] The Combahee River Collective. 1977. A Black Feminist Statement.

[32] Tina Vasquez. 2014. It's time to end the long history of feminism failing transgender women. *Bitch Media*, 17.

[33] Iris Marion Young. 2014. Five faces of oppression. *Rethinking Power* (2014), 174-95.

[34] Avital Shulner Tal et al. 2019. "End to End" Towards a Framework for Reducing Biases and Promoting Transparency of Algorithmic Systems. 2019 14th International Workshop on Semantic and Social Media Adaptation and Personalization (SMAP) (2019). DOI:http://dx.doi.org/10.1109/smap.2019.8864914

[35] Joshua R. Loftus, Chris Russell, Matt J. Kusner, and Ricardo Silva. 2018. Causal reasoning for algorithmic fairness. arXiv:1805.05859v1. Retrieved from https://arxiv.org/pdf/1805.05859.pdf.

[36] Abeba Birhane and Fred Cummins. 2019. Algorithmic injustices: Towards a relational ethics. arXiv preprint arXiv:1912.07376. Retrieved from https://arxiv.org/abs/1912.07376

[37] Sasha Costanza-Chock. 2018. Design Justice: towards an intersectional feminist framework for design theory and practice. *Proceedings of the Design Research Society* (2018).

[38] Donna Haraway. 2013. Simians, cyborgs, and women: The reinvention of nature. Routledge, New York, NY.

[39] Julia Dressel and Hany Farid. 2018. The accuracy, fairness, and limits of predicting recidivism. *Science Advances* 4, 1, eaao5580. https://advances.sciencemag.org/content/advances/4/1/eaao5580.full.pdf

[40] Os Keyes, Nikki Stevens, and Jacqueline Wernimont. 2019. The government is using the most vulnerable people to test facial recognition software. *Slate Magazine*, 17.

[41] Lelia Marie Hampton. 2020. Twitter's Preview Feature is Both Racist and Colorist. (December 2020). Retrieved from January 10, 2021 from https://leliahampton.medium.com/twitters-picture-preview-feature-is-both-racist-and-colorist-3f2be80db079

[42] Ziad Obermeyer, Brian Powers, Christine Vogeli, and Sendhil Mullainathan. 2019. Dissecting racial bias in an algorithm used to manage the health of populations. *Science* 366 (2019), 6464, 447-453.

[43] Rediet Abebe, Solon Barocas, Jon Kleinberg, Karen Levy, Manish Raghavan, and David G. Robinson. 2020. Roles for computing in social change. *In Proceedings of the 2020 Conference on Fairness, Accountability, and Transparency (FAT* '20)*. Association for Computing Machinery, New York, NY, USA, 252–260. DOI:https://doi.org/10.1145/3351095.3372871